\documentclass[12pt,english]{article}

\usepackage{ucs}
\usepackage[T2A]{fontenc}
\usepackage[utf8x]{inputenc}
\usepackage{babel}
\usepackage{graphicx}
\usepackage[small]{caption2}
\usepackage[psamsfonts]{amssymb}
\usepackage{amsmath}

\def\apgt{\ {\raise-.5ex\hbox{$\buildrel>\over\sim$}}\ }
\def\aplt{\ {\raise-.5ex\hbox{$\buildrel<\over\sim$}}\ }

\newcommand{\rs}{\mbox {$R_{\odot}$}}

\newcommand{\ms}{\mbox {$M_{\odot}$}}
\newcommand{\ace}{\mbox {$\alpha_{ce}$}}
\newcommand{\md}{\mbox {$\dot{M}$}}

\newcommand{\ls}{\mbox {$L_{\odot}$}}
\newcommand{\mch}{\mbox {$M_{Ch}$}}

\newcommand{\myr}{\mbox {~$\rm M_{\odot}$~yr$^{-1}$}}
\newcommand{\porb}{\mbox {$P_{\rm orb}$}}

\newcommand{\al}{\mbox {$\alpha_{ce} \lambda$}}

\newcommand{\pyr}{\mbox {~${\rm yr^{-1}}$}}

\def\etal{{et al.}}

\def\apj{Astrophys. J. }

\def\aap{Astron. Astrophys. }

\def\sval{\ref@jnl{Sov.~Astron. Lett.}}

\def\mn{Mon. Not. Roy. Astron. Soc.}

\def\sna{SN Ia}

\begin{document}
\pagenumbering{arabic}

\begin{center}
\textbf{Evolution of the number of accreting white dwarfs with shell nuclear burning\\
and of occurrence rate of \sna \\[32pt]}

L.R. YUNGELSON$ ^*$

\textit{Institute of Astronomy, Moscow, Russia} \\[6pt]

\end{center}

\sloppypar
\vspace{2mm}
\noindent

{\bf Abstract.}
We analyze temporal evolution of the number of accreting white dwarfs with shell hydrogen burning 
in semidetached and detached binaries. We consider a stellar system in which star formation lasts for 
10~Gyr with a constant rate, as well as a system in which the same amount of stars is formed in a 
single burst lasting for 1~Gyr.  Evolution of the number of white dwarfs is confronted to the 
evolution of occurrence rate of events that usually are identified with SN~Ia or accretion-induced 
collapses, i. e. with accumulation of Chandrasekhar mass by a white dwarf or a merger of a pair of CO 
white dwarfs 
with total mass not lower than the Chandrasekhar one. In the systems with a burst of star 
formation, at  $t=10^{10}$\,yr, observed supersoft X-ray sources, most probably, are not precursors of SN~Ia.  
The same is true for an overwhelming majority of the sources in the systems with constant star 
formation rate.
In the systems of both kinds mergers of white dwarfs is the dominant SN~Ia scenario. In symbiotic binaries, 
accreting CO-dwarfs do not accumulate enough mass for SN~Ia explosion, while ONeMg-dwarfs finish their evolution by an accretion-induced collapse with formation of a neutron star. 
  
\vfill
\noindent
{$ ^*$}email: lry@inasan.ru

\newpage
\section*{Introduction}
\label{sec:intro}
The nature of progenitors of \sna\ is still unclear. There exists a consensus that exploding objects are white dwarfs (WD) of Chandrasekhar (\mch) or larger mass, but the way of matter accumulation 
is still discussed. Historically, three main scenarios were suggested: (i) accretion from the wind 
in a symbiotic binary (Truran and Cameron, 1971), (ii) accretion in a semidetached binary (``Single-degenerate scenario'' (SD), 
Whelan and Iben, 1973), (iii) merger of components in a binary WD with total mass $\apgt \mch$ (``Double-degenerate scenario'' (DD), Webbink, 1984; Iben and Tutukov, 1984).  
Knowledge of \sna\  scenarios is of immense importance for cosmology, since in scenarios (i) and (ii) the mass of exploding object is \mch\ and the supernova may, most probably, be considered as a standard candle. 
In scenario (iii), after several tens of Myr after beginning of star formation first explode merging pairs 
with total mass sufficiently larger than \mch, while explosions of the  pairs with $M_{tot}$ close to \mch\   
start only $\sim10^9$\,yr later (Tutukov and Yungelson, 1994; Bogomazov and Tutukov, 2009).

In scenarios (i) and (ii) the binary passes through a stage in which the first-formed WD accretes the matter from 
companion. It is assumed that the (quasi)stationary hydrogen and helium burning  at the surface of 
a WD  allows to accumulate \mch. In scenario (3), a similar stage precedes formation of the second WD 
in the system. 

Following van den Heuvel et al. (1992), WD with (quasi)stationary burning at the surface are identified with  supersoft X-ray sources (SSS).
Recently, Di Stefano (2010a,b) called attention to the fact that in the galaxies of different morphological types 
there exists a significant (up to 2 orders of magnitude)
deficit of SSS as compared with expectations based on the \sna\ rate. Independently, 
to the same conclusion came Gilfanov and Bogd\'{a}n (2010), based on the study of the luminosity of 
elliptical galaxies in the supersoft X-ray range. 

In this note we study evolution of the numbers  of accreting WD and potential \sna\ progenitors. 
Computations were carried out for two cases of star formation history: (i) assuming a constant star formation rate for $10^{10}$\,yr and (ii) for a case when the same amount of mass as in the previous case is converted into stars in $10^9$\,yr. Thus, our computations exemplify \textit{idealized}
spiral and elliptical galaxies.
It is shown that at the age of 
10\,Gyr all systems with accreting WD in elliptical galaxies and their overwhelming
majority in spiral galaxies are not precursors of \sna. 
We confirm suggestion that the merger of WD is the main channel for \sna. We show that in symbiotic binaries, most probably, only oxygen-neon-magnesium WD are able to accumulate \mch; then they collapse with formation of neutron stars and do not produce \sna\footnote{We make a caveat that we use the term ``\sna'' for situations when carbon-oxygen WD accumulates \mch\ despite the fact that, strictly speaking, 
model computations of explosions of WD are still unable to reproduce all features of observed \sna.}.

\section*{The method of computations }
\label{sec:method}

Earlier, we computed populations of accreting WD, for instance, in our studies of symbiotic stars
(Yungelson \etal, 1995; L\"{u} \etal, 2006), supersoft X-ray sources (Yungelson \etal, 1996; Fedorova \etal, 2004), distributions of supernovae over redshift (Yungelson and Livio, 1998, 2000).  
The main assumptions of our population synthesis code are described in the above-mentioned papers and in Tutukov and Yungelson  (2002)\footnote{The only exception is the paper by L\"{u} \etal\ (2006)
in which we used modified by us Hurley \etal\ (2002) code.}. 
Let note only several details.

Like in our previous studies it is assumed that all stars are born in binaries and star formation rate is defined as (in yr$^{-1}$)
\begin{equation}
\label{eq:brate}
\frac{dN}{dt} = 0.2 d (\log a) \frac{dM_1}{M_1^{2.5}} f(q) dq,
\end{equation}   
where $M_1$\ is the mass of the primary component, 
$f(q)$   is the distribution of stars over mass-ratio of components $q=M_2/M_1$
normalized to 1. Distribution $f(q)$ is
\begin{equation}
\label{eq:fq}
f(q)=\begin{cases}
1                & \text{for close binary stars,} \cr
0.1055 q^{-2.5} & \text{for wide binaries with}\ 1\geq q \geq     0.3, \cr
2.14 &  \text{for wide binaries with}\ 0.3 > q \geq 0. \cr 
\end{cases}
\end{equation}

Star formation rate is normalized to the formation of one binary with 
$M_1 \geq 0.8\,\ms$\ per yr.
The range of semi-major axes of orbits is  
$6 (M_1/\ms)^{1/3} \leq a/\rs \leq 10^6$ (Kraicheva \etal, 1981; Vereshchagin \etal,  1988). 
Single stars in the model are merger products or components of binaries disrupted by supernovae explosions. For the minimum mass of components of binaries  
0.1\,\ms,  Eqs. (\ref{eq:brate}) and (\ref{eq:fq}) give model star formation rate   
$\simeq 8$\,\ms\ per yr. This value is consistent with the observational estimates of the current star formation
rate in the Galactic disk (Gilmore, 2001).
Assuming that the age of model galaxies is 
$10^{10}$\,yr, for the constant rate of star formation one obtains that the total mass of stars processed in the stars is close to $810^{10}$\,\ms, comparable to the  observational estimates
of the mass of Galactic disk (Dehnen and Binney, 1998).  The mass of model galaxies and formation rates and numbers of
different systems rendered below may be rescaled by a simple change of numerical coefficient in 
Eq.~\ref{eq:brate}.

Taking into account the specifics of the evolution of close binaries,  we assume in the model that 
the minimum mass of the components of close binaries that explode as core-collapse supernovae  is 11.5\,\ms; in wide systems this threshold is 10\,\ms. Then the occurrence rate of SN~II and SN~b,c is close to 1 per 40 yr. This estimate does not contradict observational estimate of the 
Galactic formation  
rate of pulsars -- (0.9 -- 1.9) per 100 yr  (Vranesevic \etal, 2004). Occurrence rate of \sna\ is discussed below. 
 
Stellar wind mass-loss was computed using Reamers' formula $\md=4\times10^{-12}\eta (R/\rs) (L/\ls) (M/\ms)^{-1}$\myr\ with variable parameter  
$\eta$ (see Yungelson \etal,\ 1995). 
For computation of stellar wind accretion rate standard Bondi-Hoyle formalism was applied (Bondi, 1952). 

In calculations of the numbers of accreting WD with surface nuclear burning we considered both objects 
with stationary and unstable burning (see Yungelson \etal 1996). We considered both possibility of accumulation of the matter and possibility of erosion of
WD due to unstable nuclear burning. Conditions for unstable burning of hydrogen were taken from Prialnik and Kovetz (1995), while for helium -- from Iben and Tutukov (1996).

Let list several differences to the paper of Yungelson and Livio (1996) which was immediately aimed at the study 
of WD with nuclear burning which were identified in that paper with supersoft X-ray sources.

(i) Following Kato and Hachisu (1994), in the present paper we assumed that in the case of accretion rate $|\md|\leq 10^{-4}\,\myr$ and larger than the upper limit of the rate 
of stationary hydrogen burning  
$\log(|\dot{M}_{max}|)  \approx -9.31+4.12M_{wd}-1.42M_{wd}^2$ (in \myr), excess of the matter leaves the system via optically thick stellar wind from WD and takes away specific angular momentum of WD. Under this assumption, the maximum mass ratio of components which still allows stable mass-exchange increases from  $q\approx0.78$ to $q\approx 1.15$. More, mass-exchange in the thermal time scale of the donor
becomes possible, while in the earlier versions of computations it was assumed that in all such systems a common envelope forms. Computational algorithm corresponding to these assumptions is described in detail by Yungelson and Livio (1998).  

(ii) We took into account stabilizing effect of relatively massive cores on dynamical stability of mass loss by red giants (Hjellming and Webbink, 1987). 

(iii) At difference to population synthesis studies mentioned above, for the estimate of the variation of separation of components in the first unstable mass-exchange episode in the systems with comparable masses of components,
we applied so-called $\gamma$-formalism, based on the balance of angular momentum (Nelemans et al., 2000).
In this case, separation of components varies only insignificantly in the course of mass exchange.
In all other cases of unstable mass-exchange Webbink's (1984) equation for common envelopes 
was applied, using fixed value of the product of common envelope parameter \ace\ and stellar structure parameter $\lambda$. Results presented below were obtained for  $\gamma$=1.5 and 
$\ace\lambda$=2,
since this combination of parameters allows to explain the characteristics of observed close binary WD   and other low-mass detached systems with WD components    (Nelemans \etal,  2001; Nelemans and Tout 2005).

The result of the above-mentioned changes in the code is that substantially larger fraction of 
binaries than computed before passes trough the stage of stationary mass exchange instead of forming common envelopes and the number of systems with subgiant donors increases by about an order of magnitude compared to the number obtained in Yungelson an
\etal, (1996). In symbiotic binaries, the number of WD with stationary burning increases by a factor
$\simeq$2.5. 

We did not consider the possibility of stabilization of mass exchange due to additional mass and momentum loss from the system due to stripping of red giant by WD wind (Hachisu et al., 1999). 
If the latter assumption is correct, one would expect the presence of a  significant amount of circumstellar matter
in the vicinity of \sna\ progenitors. However, existing observational data indicate the presence of hydrogen in the 
vicinity of only one of the known  ``normal'' \sna\ -- SN~2006X   (Patat \etal, 2007).  
Energy release due to collision of the supernova ejectum with its companion star should produce
a conspicuous kink in the early light curve lasting for $\sim$day following the explosion, especially notable in UV. Existing observational data apparently already now allow to exclude 
red-giant companions with   
$\sim\,1$\,\ms\ and distances between components $\sim\,10^{13}$\,cm (Kasen, 2010).
More, for the desired effect on the \sna\ rate, the efficiency of the stripping, which is a parameter of the problem, needs to have an unlikely value very close to 1.

We did not consider possible effect of magnetic fields of WD on the accretion regime, which, most probably, only diminishes
the efficiency of accretion (by about 30\%, Lipunov and Postnov (1988)).

\section*{Results and discussion}
\label{sec:results}

In the present paper, we did not aim at a detailed study of relation between the number of accreting WD 
and \sna\ for numerous variations of the parameters of population synthesis, but limited ourselves by
presenting an illustratory model which, to our opinion, clearly enough exemplifies this relation.  

\begin{center}
\begin{figure*}[t!]
\begin{minipage}[t]{0.49\textwidth}
\includegraphics[scale=0.47]{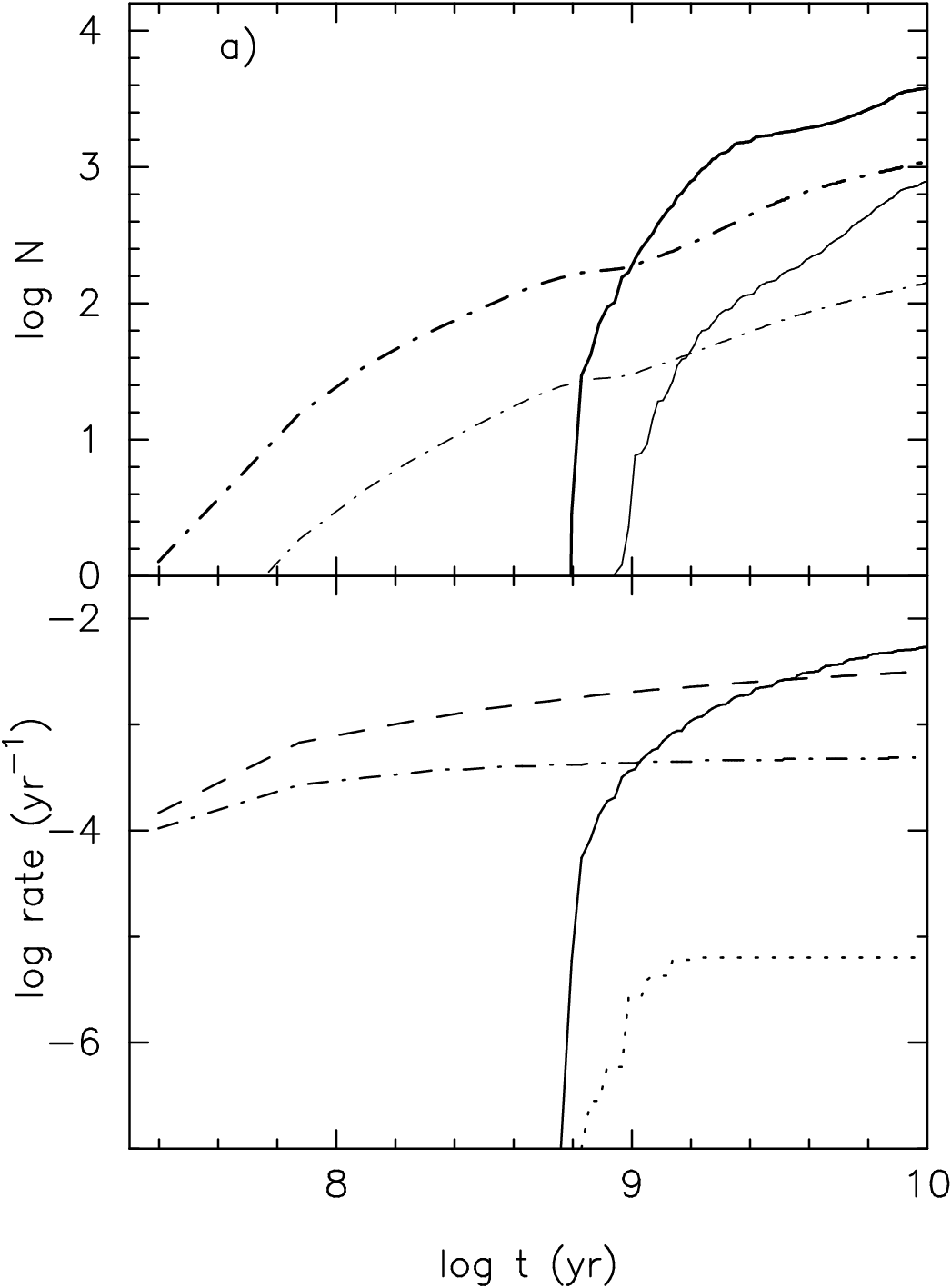}
\end{minipage}
\hfill
\begin{minipage}[t]{0.49\textwidth}
\includegraphics[scale=0.47]{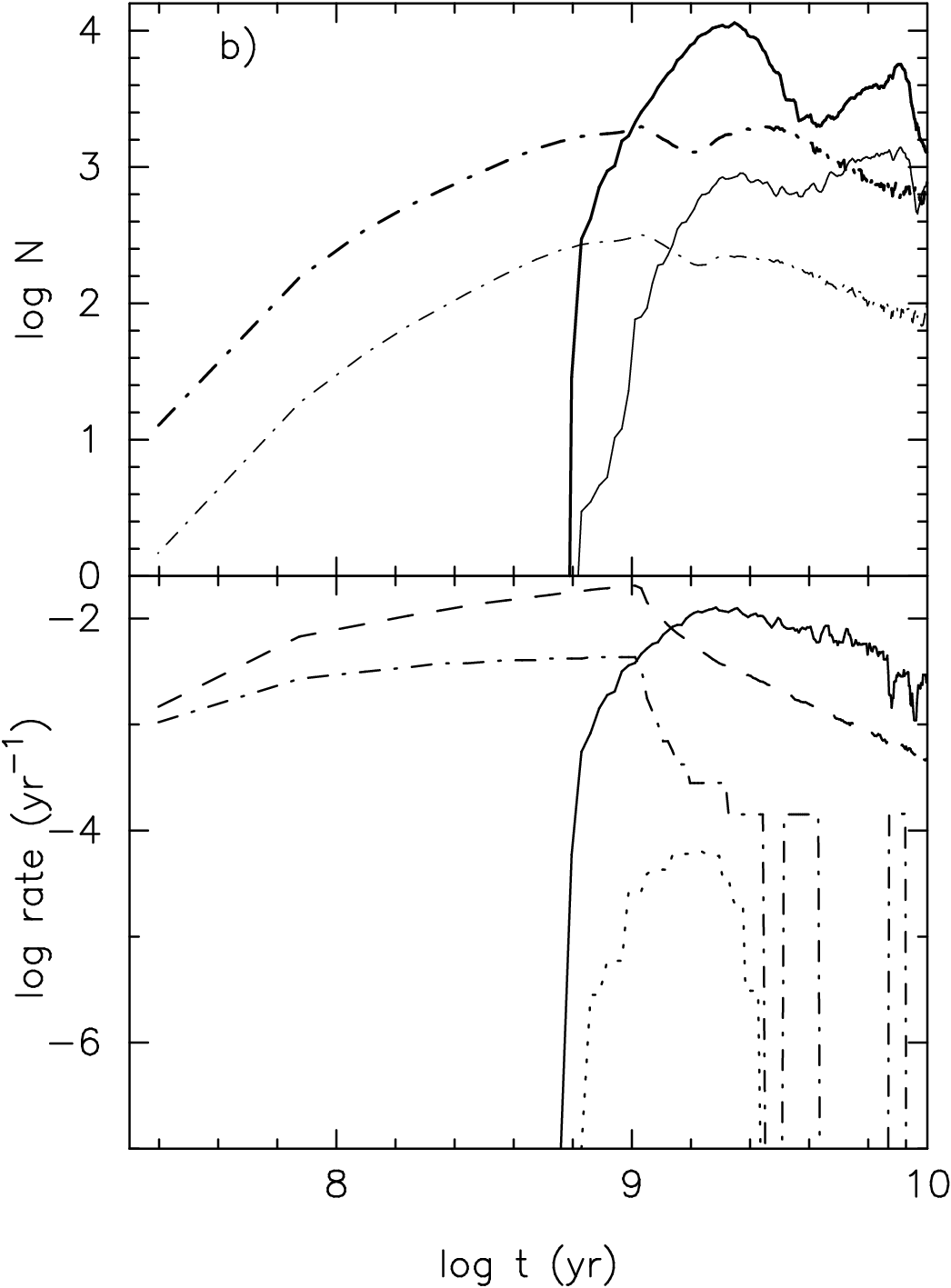}
\end{minipage}
\caption{
Evolution of the formation rate of accreting WD with surface hydrogen burning, of occurrence rate of
\sna\ and accretion-induced collapses and of the number of binaries associated with them.
а) -- a system with star formation rate constant for 
$10^{10}$\,yr. In the lower panel dashed line shows occurrence rate of \sna\ in double-degenerate scenario,
dotted line shows occurrence rate of \sna\ in single-degenerate scenario. 
Dash-dot line shows the rate of AICs in symbiotic systems, solid line -- the rate of formation of semidetached systems with hydrogen-burning WD. In the upper panel thick solid line shows the number of semidetached systems with subgiant donors in which WD burn hydrogen stationary, thin solid line -- the number of the system of the same type with WD burning
hydrogen unstably. Thick and thin dash-dot lines show the numbers of symbiotic stars with stationary and non-stationary hydrogen-burning
WD, respectively. 
b) -- the same as in (a) but for a system with a $10^9$\,yr long star-formation burst that produced the same mass of stars as in case (a).    \hfill}
\label{fig:tscales}
\end{figure*}
\end{center}

Results of computations for two model galaxies are presented in Fig.~\ref{fig:tscales}. 

\vskip 2mm
\textit{\textbf{Close binary white dwarfs and semidetached binaries }}

Yungelson \etal (1996) have shown that the population of semidetached systems with hydrogen-burning accreting WD is dominated by the systems with subgiant donors. 
Therefore, we restricted ourselves by consideration of these systems only. 

In star-formation model (i) at the age $t=10^{10}$\,yr occurrence rate of the merger of WD with total mass 
exceeding \mch\ is  $3.2\times10^{-3}$\pyr\ in agreement with the rate of \sna\ in 
Sbc-type galaxies (Cappellaro \etal, 1999). 
In case (ii) this rate is $4.5\times10^{-4}$\,\pyr.
Thus, if merger of WD is identified with \sna, the ratio of events in two cases per unit mass 
is $\simeq\,7$. This is close to the ratio $\aplt\,5$\ for E/S0 and Sbc/d galaxies found by Mannucci \etal\ (2005). More fine tuning of this ratio makes no sense for illustrative calculation. 
In case (i) occurrence rate of accumulations of \mch\ in semidetached systems at the age  
$t=10^{10}$\,   
is  $6\times10^{-6}$\,\pyr. Qualitatively, the ratio of mergers and \mch\ accumulations does not differ from the ratio found in our earlier
computations, where it never exceeded  $\sim 0.1$. The main reason for the low efficiency of the 
SD-scenario is, in our case, low efficiency of matter accumulation: 
combinations of masses of WD and accretion rates is favorable for unstable hydrogen burning and loss of 
most of accreted hydrogen in Novae explosions.

Formation rate of close binary WD reflects SFR, the mergers start several tens of Myr after beginning of star formation, when the binaries that had ZAMS masses of
components close to 10\,\ms\ start to form the pairs of WD. 
More specifically, time-delay of \sna\ in DD-scenario depends on metallicity of stars which defines 
formation time of WD and their mass (see, for instance, Kobayashi \etal\ (2000)) and on the parameter of common envelopes, since, in 
crude approximation $a_f \propto \al a_0$, where $a_0$ and $a_f$ are initial and final separations of components. 
In $10^{10}$\,yr merge the pairs of WD with $a_f \sim\,\rs$.
In the spiral galaxies new pairs of WD form continuously and merger rate continuously increases, since merge both ``old'' initially relatively wide pairs and ``young'' relatively close ones. 
In elliptical galaxies a reservoir of pairs with different combinations of masses of components 
and separations is created during the burst of star formation and it  gradually ``melts'' and therefore at certain moment after cessation of star formation merger rate starts to decline. 

Formation of an accreting CO-WD with initial mass lower than 1.2\,\ms\ and 
accumulation of \mch\ by a dwarf 
require certain time, therefore in semidetached systems \sna\ occur with delay  
of $\sim\,10^{9}$\,yr respective to star formation. Numerous studies of delay time distributions for \sna,
starting from the papers by Jorgensen \etal (1997) and Yungelson and Livio (2000),
show that after a single star-formation burst semidetached systems in which a CO-WD is able to accumulate 
\mch\ form and exist over a limited span of time which lasts for several Gyr. In our model, these are systems
in which at the instant of Roche-lobe overflow by the donor $M_{wd,0} \apgt 0.85$\,\ms\ and  
$M_{sg,0} \apgt 1.4$\,\ms.  
Figure~\ref{fig:tscales} shows, 
that formation rate of semidetached systems with WD able to accumulate \mch\ even in relatively early stages of galaxy evolution is only a minor fraction of the total formation rate of semidetached systems 
with WD with surface hydrogen-burning. 
Semidetached systems with WD form during total lifetime of galaxies evolution, but 
\textit{in the case of burst-like star formation the systems observed at present are not precursors of \sna}. Currently, in the model galaxy with continuous star formation  \sna\ occur in semidetached systems formed approximately  
$6 \times 10^8$ to $2.5 \times  10^9$\ yrs ago. 

Along to the systems in  which hydrogen burns stationary, there exist systems in which hydrogen at the surface of WD
burns in outbursts. The systems of two types are shown in Fig.~\ref{fig:tscales} separately.
For the outbursting systems, we assumed, like in Yungelson \etal\ (1996) that hydrogen burning time is equal to the time of decline of bolometric luminosity of WD by 3 stellar magnitudes 
($t_{3bol}$). The value of the latter was estimated by interpolation in Prialnik and Kovetz (1995) data. 
Both systems with stable and unstable hydrogen burning start to form almost simultaneously, but at any time the former dominate, since     
$t_{3bol}$\ is short. In the systems with subgiants the fraction of outbursting systems is initially low, since first start to overfill their Roche lobes relatively massive stars with high mass-loss rates. Later, the fraction of systems in which the rate of accretion is lower than the limit for stable
hydrogen burning increases since the systems with massive donors finish their evolution fast.

\vskip 2mm
\textit{\textbf{Symbiotic stars}}

The second main group of stars with accreting WD are symbiotic stars in which WD are accompanied by giants or supergiants. White dwarfs accrete from the wind of companion.
In the initial stage of the evolution of donor along the giant branch stellar wind is weak. In wide systems
with orbital periods of hundreds and thousands day accretion initially is extremely inefficient.
It takes a long time to accumulate enough hydrogen for the first outburst. 
Then WD transits into  regime of unstable burning and, typically only shortly before the loss of envelope 
by the giant WD may become a stationary burner (see for details Yungelson \etal\ (1995) and L\"{u}
\etal\ (2006)). Conventionally, we shall denote as ``symbiotic stars'' all detached systems in which
at least one outburst of nuclear burning at the surface of WD had happened, irrespective to whether such stars 
correspond to formal phenomenological definition of symbiotic stars.  

Only $\sim 0.01\%$\ of all systems of giants and supergiants with WD companions reach the stage of 
symbiotic star. Duration of the stage of stationary burning of hydrogen at the surface of WD is, as a rule,
only $\sim10^5$\,yr. At $t=10^{10}$\,yr the ``spiral'' and ``elliptical'' model galaxies harbor
1360 and 1260 systems with hydrogen-burning WD, respectively. Currently, about 200 symbiotic stars are known in the Galaxy (Belczynski \etal, 2000). 
Observational estimates based on extrapolation of the number of known objects and estimates of completeness of the observed sample range from   
$3\times10^3$ (Allen, 1984) to $3\times10^5$ (Munari and Renzini, 1992) objects. 
As a result of a pointed survey which took into account selection effects, Corradi \etal\ (2008, 2010) found only 11 new symbiotic star in the sky region where 11 stars were already known before. Thus, it seems that expectations of existence of hundreds of thousands of symbiotic stars in the Galaxy are 
exaggerated and our estimate of 
$\sim 1000$\ objects is realistic.
Time delay of formation of symbiotic stars with respect to star formation
is several tens of Myr. This is slightly more than the time of formation of the first WD. 
Symbiotic stars continue to form for  10~Gyr both in spiral and elliptical galaxies. In the latter case this happens as main-sequence is left by less and less massive stars.  
As a result, at 
$t=10^{10}$\,yr the numbers of WD accreting from the wind in the model ``spiral'' and ``elliptical'' 
galaxies are comparable.    

\begin{center}  
\begin{figure*}[t!]
\includegraphics[angle=-90,scale=0.5]{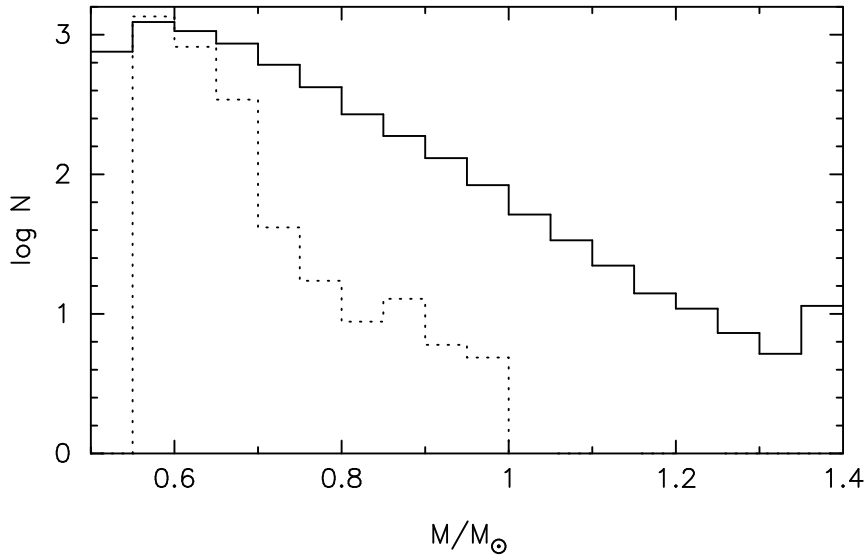}
\caption{Mass spectrum of accreting WD with nuclear burning at the surface. Solid line -- WD in the 
model system with continuous star formation, dotted line -- WD in the model system with burst-like star formation. 
In both cases we show the sum of the numbers of WD with stationary and nonstationary burning.   
\hfill}
\label{fig:sssmass}
\end{figure*}
\end{center}

Among symbiotic stars about 2/3 of systems  are wide, precursors of WD in them did not overflow their Roche lobes. Therefore, mass spectrum of WD in symbiotic stars is dominated by low-mass 
CO WD 
($\sim\,0.6$\,\ms). Since accretion rates are low, nuclear burning is mostly unstable and  in most cases erosion of white dwarfs happens instead of accumulation of mass.  Estimates by L\"{u} \etal\
show that in symbiotic stars during accretion stage mass of WD increases by no more than 
$\simeq 0.1$\,\ms.
 This has an important consequence. The masses of new-born CO WD do not exceed 
(1.1 -- 1.2)\,\ms\ (Nomoto, 1984; Ritossa \etal, 1996; Poelarends \etal, 2008). More massive WD are composed by oxygen, neon, and magnesium. Hence, in symbiotic stars CO WD are unable to accumulate
\mch. Instead of \sna\ accretion-induced collapses (AIC) of ONeMg-WD occur which prduce neutron stars 
(Nomoto and Kondo,  1991). The rate of AIC's in the model of a spiral galaxy 
$6.6\times 10^{-4}\pyr$  is consistent with the estimate of AIC's in the Galaxy based on the yields 
of $r$-process elements, if one takes into account 
that, according to 2-D rotating models of AIC (Fryer \etal, 1999; Dessart \etal, 2006) only 
$\sim 0.001$\,\ms\ of matter enriched by $r$-process elements is ejected during AIC.  
In this case, evolution of symbiotic stars results in formation of wide binaries composed by neutron stars and WD.  
AIC's, themselves, probably explain the weakest peculiar \sna\ (Metzger \etal, 2009).

Since binaries with more massive WD complete their evolution faster, 
in the elliptic galaxies their number must be lower than in spiral ones.  This is clearly shown by Fig.~\ref{fig:sssmass}, in which we render mass spectra of
accreting WD in the model galaxies with different star formation history. This circumstance has to influence the number of detectable WD with nuclear burning.
 

\section*{Conclusion}
\label{sec:concl}

Above, we presented results of a computation that exemplifies evolution of the number of accreting WD in semidetached binaries  with subgiant donors and symbiotic stars  and evolution of the \sna\ rate in the galaxies with continuous and burst-like star formation. The following conclusions may be drawn. 

1. In both models of galaxies, after 10~Gyr of evolution WD with surface hydrogen burning exist in approximately equal numbers (several thousands, Fig.~\ref{fig:tscales}). Test computations show that this result does not 
change qualitatively if, for instance, other parameters of common envelopes 
(\al=0.5 or 1), different parameterization of  mass-loss rate, Tutukov and Yungelson (1979)
equation for common envelopes, different efficiency of matter accumulation are assumed. The number of objects and the rate of \sna\ vary within 30\%\footnote{Let note only that for \al=0.5\ 
at $t=10^{10}$\,yr\ in the model of elliptical galaxy \sna\ are absent at all.}.
Using, for instance, a star formation function with an  initial
burst and a tail extending for several Gyr in the model of elliptical galaxy also would not change results, since,
because of short delay times all ``old'' \sna\ in semidetached systems succeed to explode in 10~Gyr.
  
Supersoft X-ray sources in elliptical galaxies are observed, but according to model computations
among semidetached systems with subgiant donors precursors of \sna\ are absent. In the model of a spiral galaxy such precursors do exist. 
Potential, but not immediate  \sna\ precursors both in spiral and elliptical galaxies may be symbiotic stars 
if the following conditions are fulfilled. Donor-star has to overfill Roche lobe in the course of evolution and 
to form a common envelope. In the common envelope separation of 
WD and the nucleus of the donor must become small enough  to enable merger in the future. The sum of the mass of already existing 
WD and the core of the donor must to exceed \mch.   
As our model computations show, for similar total masses of galaxies, at the age of 10~Gyr the rate 
of \sna\ in model elliptical galaxy is by 2 orders of magnitude lower than in the spiral one.   

2. Every of relatively well studied nearby spiral galaxies 
(M31, M101, М51, M83) contains about 100 SSS, in nearby elliptical galaxies (NGC4697, NGC4472)
their number is 
several dozens  (Di Stefano, 2010a).  Taking into account that, according to the estimates of Di Stefano, \textit{Chandra}
is able to observe in these spiral galaxies only objects with mass $\apgt (1.0 - 1.2)$\,\ms\ and 
objects with $M \apgt 0.8$\,\ms in elliptical galaxies, the number of observed SSS agrees with expected from observations (Fig.~\ref{fig:sssmass}).

3. The main scenario for \sna\ in our model is merger of CO WD. In the model spiral galaxy  
occurrence rate  of accumulations of \mch\ by WD in semidetached systems with subgiant donors 
at $t=10^{10}$\,yr is by two orders of magnitude lower than merger rate. In the model elliptical galaxy merger becomes a sole mechanism of \sna\ in  $\simeq 2.5\times10^9$\,yr\ since beginning of star formation. A typical time scale of single-degenerate \sna\ ``epoch'' of the order of 
$10^9$\,yr is found in practically all studies of delay time distributions of \sna\ (see, e.g., Ruiter \etal, (2009), Mennekens \etal, (2010) and references therein). For explanation of \sna\ delays 
up to  $\sim 10^{10}$\,yr in SD-scenario it is necessary to assume stabilization of mass-exchange by reemission from the system of a significant fraction  of the matter lost by the donor and low 
specific angular momentum of ejected matter.   
As a counterargument against this hypothesis may be used the fact that the traces of hydrogen are 
discovered as yet for only one \sna. Even more, almost commonly accepted assumption of possibility of ejection of a significant amount of matter by optically thick wind is put in doubt (Badenes
\etal, 2007): the wind should result in formation of cavities in ISM around supernovae remnants which are not observed. 
If assumption about ejection of excess of matter is incorrect, real occurrence rate of \sna\ in semi-detached systems may be even
lower than our estimates. 

Despite our model differs from the models of other authors (e.g. Ruiter \etal, (2009)) by lower efficiency of accumulation of the matter by WD, qualitative conclusions are similar: most probably, in spiral galaxies both DD- and SD-scenarios work for their total lifetime, but DD-scenario dominates, while in elliptical galaxies only DD-scenario remains efficient in  several Gyr after cessation of star formation.
Thus, we confirm conclusions made by Jorgensen \etal, (1997), Yungelson and Livio (2000), Mennekens \etal, (2010).  

Let note that lately several close binary WD or their immediate precursors which have 
$(M_1+M_2) \simeq \mch$ and will merge in less than Hubble time were found. These are binary 
WD~2020-425 (\porb=0.3\,day, $M_1+M_2=1.348\pm0.045$\,\ms, 
Napiwotzki \etal, 2007), 
a subdwarf accompanied by a white dwarf KPD~1930+2752 (\porb= 0.37765$\pm$0.00002 day,   
$M_1+M_2=(1.36 − 1.48)$\,\ms, Geier \etal,  2007), 
planetary nebulae nuclei with WD companions TS~01 (\porb=0.163~day,
$M_1 =0.54\pm 0.02$\,\ms,  $M_2\approx 0.86$\,\ms, Tovmassian \etal,  2010) and 
V458~Vul ($\porb \approx 0.068$~day,
$M_1\approx0.6$\,\ms,  $M_2\apgt 1.0$\,\ms, Rodr{\'{\i}}guez-Gil \etal,  2010).

In the process of CO WD merger, less massive WD transforms into a disk-like structure around more massive companion. It was usually assumed that in this configuration temperature maximum is located at the ``disk-dwarf'' interface and that carbon burning starts there. In this model burning front propagates inward and transforms a CO WD into an ONeMg one, which collapses with formation of a neutron star. However, Yoon \etal, (2007) have shown that under certain conditions the rate of settling of matter from disk onto WD may be low enough for commencing of burning in the center of WD and a \sna. The model still must be elaborated but it is quite possible that ``theoretical'' 
objections against DD-scenario may be removed. As an additional argument in favor of DD-scenario one may consider discovery of several \sna\ for which the estimates of the mass of ejected radioactive Ni are close to \mch\ or even exceed it:  
SN~2003fg  (Howell \etal, 2007), 
SN~2006gz (Hicken \etal, 2007),
SN~2007if (Scalzo \etal, 2010; Yuan \etal, 2010), 
SN~2009dc (Silverman \etal, 2010).  

4.  In the systems with WD accreting from stellar wind, efficiency of matter accumulation by WD is very low. In these systems, according to our computations, only ONeMg WD can accumulate \mch. We expect that in such systems evolution  should end not by \sna, but by accretion-induced collapse and formation of neutron star+WD pairs.

\vskip 3mm

The author acknowledges fruitful discussions with N.N. Chugai, K.A. Postnov, and M.R. Gilfanov.
This study was partially supported by RFBR grant 10-02-00231 and by the Program of the Presidium of the Russian Academy of Sciences ``Origin and evolution of stars and galaxies''.
 
\section*{References}
\label{sec:refs}
\small{\begin{enumerate}
\item D.A. Allen, Pub. Astron. Soc. Austral. {\bf 5}, 369 (1984).
\item  C. Badenes, J.P. Hughes, E. Bravo, \etal,   Astrophys. J. {\bf 662}, 472 (2007).
\item K. Belczy{\'n}ski, J. Miko{\l}ajewska, J., U. Munari, \etal,
 Astron. Astrophys. Suppl. Ser.  {\bf 146}, 407 (2000).
\item A.~I. Bogomazov, A.~V. Tutukov, Astron. Rep, {\bf 53}, 214 (2009)
\item H. Bondi, \mn\  {\bf 112}, 195 (1952).
\item  E. Cappellaro, R.  Evans, R. and М. Turatto,
\aap {\bf 351}, 459 (1999).
\item R.L.M. Corradi, E.R. Rodr{\'{\i}}guez-Flores, A. Mampaso, \etal,
\aap    {\bf 480}, 409 (2008).
\item R.L.M. Corradi, M. Valentini, U. Munari, \etal,
\aap    {\bf 509}, A41 (2010).
\item W. Dehnen and J. Binney, \mn\ {\bf 294}, 429 (1998). 
\item  L. Dessart, A. Burrows, C. Ott, \etal, \apj\ {\bf 664}, 1063 (2006).
\item R. Di Stefano, Astrophys. J. {\bf 712}, 728 (2010а).
\item R. Di Stefano, arXiv:1004.1193 (2010b). 
\item A.V. Fedorova, A.V. Tutukov, L.R. Yungelson, Astron. Lett {\bf 30}, 73 (2004). 
\item  C.L, Fryer, S.E. Woosley, M. Herant, M., \etal,   Astrophys. J. 
{\bf 520}, 650 (1999).
\item S. Geier, S. Nesslinger, U. Heber, \etal,   Astron. Astrophys.  
{\bf 464}, 299 (2007).
\item M. Gilfanov and A. Bogd\'an, Nature {\bf 463}, 924 (2010).
\item G. Gilmore, ASPC, {\bf 230}, 3, 2001.
\item I. Hachisu, M. Kato and K. Nomoto,  Astrophys. J. {\bf 522}, 487 (1999).
\item  M. Hicken, P.M. Garnavich, J.L. Prieto, \etal,  Astrophys. J. 
{\bf 669}, L17 (2007).
\item M.S. Hjellming and R.F. Webbink,  Astrophys. J. {\bf 318}, 794 (1987).
\item  D.A. Howell, M. Sullivan, P.E. Nugent, \etal, Nature {\bf 443}, 308 (2006). 
\item J.R. Hurley, C.A. Tout and O.R. Pols,  \mn {\bf 329}, 897 (2002).
 \item  I. Iben Jr. and A.V. Tutukov,  Astrophys. J. Suppl. Ser.  
{\bf 54},  335 (1984).
\item  I. Iben Jr.and  A.V. Tutukov, Astrophys. J. Suppl. Ser. 
{\bf 105}, 145 (1996). 
\item  H.E. Jorgensen, V.M. Lipunov, I.E. Panchenko, \etal, 
Astrophys. J. {\bf 486}, 110 (1997).
\item M. Kato and  I. Hachisu, Astrophys. J. {\bf 437}, 802 (1994).  
\item  D. Kasen,  Astrophys. J. {\bf 708}, 1025 (2010).
\item C. Kobayashi, T. Tsujimoto and K. Nomoto, Astrophys. J. {\bf 539}, 26  (2000). 
\item Z.~T. Kraicheva, E.~I. Popova, A.~V. Tutukov 
A.~V., \etal, SvAL, {\bf 7}, 269 (1981) 
\item  V.M. Lipunov and  K.A. Postnov, Astrophys. Space. Sci. {\bf 145}, 1 
 (1988). 
\item   G. L\"{u}, L. Yungelson and Z. Han, \mn\ {\bf 372}, 1389 (2006).
\item F. Mannucci, M. Della Valle, N. Panagia, \etal,
Astron. Astrophys.  {\bf 433}, 807 (2005).
\item  N. Mennekens, D. Vanbeveren, J.-P. De~Greve \etal,
\aap 515, A89 (2010).
\item B.D. Metzger, A.L. Piro, and E.Quataert, \mn\ {\bf 396}, 1659 (2009).  
\item  U.Munari, A. Renzini, Astrophys. J. {\bf 397}, L87  (1992). 
\item R. Napiwotzki, C.A. Karl, G. Nelemans, \etal, ASP Conf. Ser. {\bf 372}, 387  (2007).
\item  G. Nelemans and C.A. Tout, \mn\ {\bf 356}, 753 (2005). 
\item G. Nelemans, F. Verbunt, L.R.  Yungelson, \etal,
 Astron. Astrophys. {\bf 360}, 1011  (2000).
\item G. Nelemans, L.R. Yungelson, S.F. Portegies Zwart, \etal, Astron. 
Astrophys. {\bf 365}, 491 (2001)
\item K. Nomoto, Astrophys. J. {\bf 277}, 791 (1984).  
\item K. Nomoto and  Y. Kondo, Astrophys. J. {\bf 367}, L19  (1991). 
\item F. Patat, P. Chandra, R. Chevalier. \etal, Science, {\bf 317}, 924 (2007). 
\item D. Prialnik and  A. Kovetz, Astrophys. J.  {\bf 445}, 789 (1995).
\item  A.J.T. Poelarends, F. Herwig, N. Langer, \etal, 
Astrophys. J.  {\bf 675}, 614 (2008).
\item A.~J. Ruiter, К. Belczynski  and С. Fryer, \apj {\bf 699}, 2026 (2009).
\item C. Ritossa, E. Garcia-Berro and I. Iben Jr., Astrophys. J. 
{\bf 460}, 489  (1996).  
\item P. Rodr{\'{\i}}guez-Gil, M. Santander-Garc{\'{\i}}a,  
	C. Knigge, \etal, arXiv1006.1075 (2010).
\item  J.M. Silverman, M. Ganeshalingam, W. Li, \etal, arXiv:1003.2417 (2010).
\item  R.A. Scalzo, G. Aldering, P. Antilogus, \etal, \apj\  {\bf 713}, 1073 (2010).
\item  G. Tovmassian, L. Yungelson, T. Rauch, \etal, Astrophys. J. {\bf 714}, 178 (2010).
\item  J.W. Truran and  A.G.W. Cameron, Astrophys. Space. Sci.
{\bf 14}, 179 (1971). 
\item A.V. Tutukov and L. Iungelson, IAU Symp. {\bf 83}, 401 (1979). 
\item  A.V. Tutukov  and L. Yungelson, \mn\ {\bf 268}, 871   (1994). 
\item  A.V. Tutukov and  L.R. Yungelson, Astron. Rep. {\bf 46}, 667  (2002).
\item J. Whelan and I. Iben Jr., Astrophys. J. {\bf 186}, 1007 (1973).
\item  E.P.J. van den Heuvel, D. Bhattacharya, K. Nomoto, \etal,  Astron. Astrophys. {\bf 262}, 97 (1992).
\item  S. Vereshchagin, А. Тутуков, Л. Юнгельсон \etal, Astrophys. Space. Sci.,   
{\bf 142}, 245 (1988). 
\item N. Vranesevic, R.N. Manchester, D.R. Lorimer \etal, \apj, {\bf 617}, L139 (2004).
\item  R.F. Webbink, Astrophys. J. {\bf 277},  355  (1984).
\item  C.-S. Yoon, P. Podsiadlowski and  S. Rosswog, \mn\ {\bf 380}, 933  (2007).
\item  F. Yuan, R.M.  Quimby, J.C. Wheeler, \etal,  arXiv:1004.3329 (2010).
\item L.R. Yungelson, M. Livio, J.W. Truran, \etal,  Astrophys.
J. {\bf 466}, 890  (1996). 
\item L.R. Yungelson, M. Livio, A. Tutukov, \etal,.  Astrophys.
J. {\bf 447}, 656  (1995).
\item L.R.  Yungelson and M. Livio,  Astrophys. J. {\bf 497}, 168 (1998).
\item L.R.  Yungelson and M. Livio, Astrophys. J.  {\bf 528}, 108 (2000).

\end{enumerate}
} 

\end{document}